\documentstyle[aps,graphicx,epsf]{revtex}

\begin{document}

\title{A structure factor approach to the Sunyaev-Zel'dovich effect}

\author{Alfredo Sandoval-Villalbazo$^{*\dag}$}

\address{~}
\address{$^*$Relativity and Cosmology Group, School of
Computer Science and Mathematics, Portsmouth University,
Portsmouth~PO1~2EG, United Kingdom}

\address{$^\dag$Departamento de Ciencias, Universidad
Iberoamericana, Col. Lomas de Santa Fe~01210, Mexico~D.F., Mexico}

\maketitle

\begin{abstract}

Structure factor theory (SFT) is applied to analyze distortions to
the Cosmic Microwave Background Radiation (CMBR) due the
Sunyaev-Zel'dovich Effect (SZE) in typical clusters of galaxies.
It is shown that a suitable structure factor associated to a
collisionless (Knudsen) gas is able to reproduce the SZE
distortion of the CMBR . Physical arguments suggest that SFT
yields an approximate solution to the Kompaneets equation and sets
plausible values for the two ``free'' parameters involved in the
calculation of the spectral distortion. Finally, further
applications of the SFT to other standard cosmological problems
are briefly discussed.

\end{abstract}

\bigskip

\textbf{I. Introduction}

\bigskip

For decades, light scattering theory has been applied to the analysis of
time-dependent fluctuations of the thermodynamical variables of simple
fluids. It is a well-known fact that the Laplace-Fourier transform of the
time correlation function of density fluctuations gives rise to a \emph{%
scattering law} that describes the general behavior of the
scattered spectra. The mathematical framework in which the
scattering law is established in different cases is called
Structure Factor Theory (SFT). Structure factors have been studied
from different points of view, mainly involving kinetic theory
\cite{Yip1} and hydrodynamics \cite{Berne}. For astrophysical
systems, low densities prevent the calculation of structure
factors in terms of hydrodynamical models, but the general form of
the appropriate scattering law for a collisionless (Knudsen) gas
is perfectly established \cite{Egelstaff}. Thus, the application
of SFT to the analysis of distortions of the Cosmic Microwave
Background Radiation (CMBR) is possible. Specifically we shall
concentrate in CMBR scattering by the hot gas present in clusters
of gases, better known as the thermal Sunyaev-Zel'dovich Effect
(SZE).

\bigskip

The SZE ~\cite{SZ1} provides a potentially very important probe of
large-scale structure and the main cosmological parameters (see
e.g.~\cite{b,Silk} ). The SZE is associated to Compton scattering
of CMBR photons by hot electrons in the ionized intra-cluster gas.
On average, photons are shifted to higher frequencies, but the
effect vanishes at a critical frequency where non-relativistic
calculations give $\sim 217$~GHz for typical gas temperatures
$\sim 10^{8}$~K (Relativistic corrections to the SZE ~\cite{rel}
introduce only a very small increase in the critical frequency).
The first approach to the SZE was made by means of the Kompaneets
diffusion equation, which describes photon diffusion in an
electron gas.

\bigskip

This paper shows that the CMBR distortions associated with the
thermal SZE can be reproduced by means of SFT. The picture is very
simple. As Rephaeli pointed out \cite{rel}, most photons are not
scattered \emph{even once} by the gas components. Thus, if we
consider a single frequency in the CMBR photons, the corresponding
line will be slightly broadened due to the scattering of some
photons. This type of broadening can be described by SFT. The
whole spectral distortion is then described by convolution-type
integrals that have already been applied in other CMBR problems
\cite{conv1}. The resulting curves describing spectral distortion
will represent solutions to the Kompaneets equation.  The
underlying physical basis of these solutions deserves further
attention.

\bigskip

This paper is divided as follows: section 2 is dedicated to the
analysis of the SFT approach to the thermal SZE in terms of the
scattering law suggested in Ref. \cite{Egelstaff}. Section 3
includes a critical comparision of the curve  obtained in section
2 and the SZE curve obtained by means of the Kompaneets equation.
Finally, ideas related to the physical interpretation of the
``free'' parameters involved in the structure factor description
of the SZE, and further applications of this approach, close this
work.

\bigskip

\textbf{II. Scattering law for a simple collisionless gases.}

\bigskip

A very simple way of deriving the scattering law for a colisionless gas \cite
{Egelstaff} starts by considering the generic form of a Maxwell-Boltzmann
distribution function of speeds for a simple, non-relativistic gas.
\begin{equation}
f\left( v\right) =A\,v^{2}e^{-\eta (v)^{2}}  \label{MB1}
\end{equation}
The most probable velocity for a given $\eta $ is:
\begin{equation}
v_{o}^{2}=\frac{1}{\eta }  \label{MB2}
\end{equation}
The probability that a particle has a velocity $\mathbf{v}$ is
proportional to $e^{-\left( \frac{v}{v_{o}}\right) ^{2}}$. The
displacement of a particle in time $\tau $ through the distance
$\mathbf{r}$ corresponds to $\mathbf{v}$ $\tau $ . Hence, the
normalized probability of finding it at  $\mathbf{r}$ at time
$\tau $, after it was at $\left( 0,0\right) $, is
\begin{equation}
G =\frac{1}{\pi ^{\frac{3}{2}}\left| v_{o}\tau \right|
^{3}}e^{-\frac{r^{2}}{\left( v_{o}\tau \right) ^{2}}}
\label{tcf1}
\end{equation}
Performing \ a Fourier transform in space and a Laplace transform
in time of Eq. (\ref{tcf1}), the scattering law for the gas
corresponds to the \emph{ structure factor}:
\begin{equation}
S(\nu)=\frac{1}{\sqrt{\pi }k\,v_{o}}e^{-\frac{\left( \nu \right)
^{2}}{\left( kv_{o}\right) ^{2}}}=\frac{1}{\sqrt{\pi }\alpha
}e^{-\frac{ \left( \nu \right) ^{2}}{\alpha ^{2}}} \label{Sk1}
\end{equation}
In a collisionless gas, particles are not interacting. It is known
that, in this case, the intensity of radiation scattered off the
fluid will be changed from its initial value by an amount
proportional to the Laplace-Fourier transform of the density time
correlation function, i.e. Eq. (\ref{Sk1}) \cite{Stanley}.
Physically, Eq.(\ref{Sk1}) represents a broadening of a single
spectral line of frequency $\upsilon $ due to the scattering of
incoming photons off the single particles of the collisionless
gas. A simple thumb rule corresponding to this broadening is:
\begin{equation}
\delta \left( \nu\right) \longrightarrow \frac{1}{\sqrt{\pi
}\alpha }e^{-\frac{ \left( \nu \right) ^{2}}{\alpha
^{2}}}=S(k\mathbf{,}\nu \mathbf{)} \label{thumb}
\end{equation}
Now, if the intensity of the incoming radiation, per unit of solid angle has
a blackbody spectrum:
\begin{equation}
I_{o}\left( \nu \right) =\frac{2h\nu
^{3}}{c}\frac{1}{e^{\frac{h\nu }{k_{B}T} }-1}  \label{bb1}
\end{equation}
then, $\ $the full distorted spectrum will be given by the convolution-type
integral:
\begin{equation}
I\left( \nu \right) =\int_{0}^{\infty }I_{o}\left( \omega \right)
S(k\mathbf{ ,}\omega-\nu  \mathbf{)}d\omega  \label{fsd}
\end{equation}
and the departures from the perfect black-body spectrum will be
given by:
\begin{equation}
\Delta I\left( \nu \right) =I\left( \nu \right) -I_{o}\left( \nu
\right) \label{Delta1}
\end{equation}
In order to show how $\Delta I\left( \nu \right) $ behaves, let us
suppose that  CMBR photons are traveling across a very hot gas,
$T=10^{8}K$. The typical broadening parameter $\alpha $ can be
estimated for low density systems (as the hot intra-cluster gas of
the SZE) by:
\begin{equation}
\alpha \sim \frac{2\pi }{c}\sqrt{\frac{k_{b}T}{m}}\nu
\label{alpha1}
\end{equation}
Where $k_{b}$ is Boltzmann's constant and $m$ is a single particle
mass. Fig. 1  shows $\Delta I\left( \nu \right) $ taking $m=$
$1.6\times 10^{-24} $ g (the proton mass).  As a reference, we
plot in Fig. 1 the SZE curve for the same temperature with Compton
parameter $y=10^{-4}$, as in \cite{Carlstrom}. More information
about this parameter will be given in the next section.  Not only
the shape of the curves are very much alike, but also the order of
magnitude for the CMBR distortion is practically the same.
\bigskip
\begin{figure}
\epsfxsize=3.4in \epsfysize=2.6in \epsffile{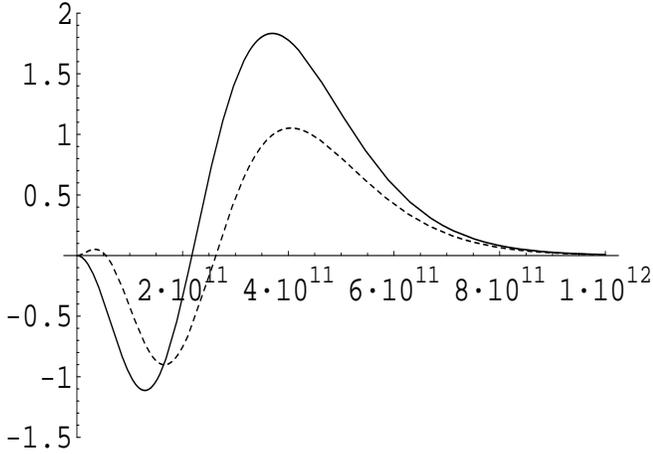}\vspace{0.5cm}
\caption{The thermal SZE standard curve (solid) compared to the
CMBR  spectrum distortion computed from SFT (dashed), $\Delta I$
$(erg\,\,s^{-1}Hz^{-1}cm^{-2}ster^{-1}\times 10^{-18}),$ against
frequency ($Hz$) for $T=10^{8}K$.}
\end{figure}
\vspace{0.5cm}
\bigskip

\textbf{III.} \textbf{The }$y$\textbf{\ Compton parameter in SFT}

\bigskip

Up to this point, it has been shown that, mathematically, SZE-type
curves can be obtained by means of the convolution integrals
(\ref{fsd}). Physically, the $\alpha $ parameter is a simple line
broadening proportional to the width parameter of the low density
system. Of course, no sound waves exist in a system such as the
hot intra-cluster gas (densities are close to $10^{-3}$ particles
per cubic centimeter), but the broadening does exist and has been
the subject of several studies \ regarding the transition from a
collision-dominated system,  to a collisionless one \cite{Yip1} .
In dense systems, sound waves will also scatter off the CMBR
photons and the structure factor would have to take into account
Brillouin scattering \cite{Berne}.  The thumb rule (\ref{thumb})
would then involve Lorentzian shapes an in the case of liquids
\cite{Stanley}.

\bigskip

Nevertheless, more physics is needed to get a better description
of the SZE curves. In particular, the scattering cross section
between photons and particles, and the density and size of the
cluster enter in the ordinary approach to the SZE by means of the
\emph{Compton ``}$y"$\emph{\ parameter}:
\begin{equation}
y=\int n\sigma \frac{m_{e}c^{2}}{k_{b}T}d\ell   \label{yc}
\end{equation}
Here $n$ is the number density of the electrons in the cluster,
$\sigma $ is the Thomson cross section and the integral is
supposed to be taken along the line of sight of a photon. Roughly,
we can write:
\begin{equation}
y\sim n\sigma \frac{m_{e}c^{2}}{k_{b}T}\,\ell   \label{yb2}
\end{equation}
In this case, $m_{e}$ is the electron mass $.$ For typical
temperatures $\left( T=10^{8}K\right) $:
\begin{equation}
y\sim 10^{-4}  \label{yb3}
\end{equation}
The actual value for $y$  depends on the precise properties of the
cluster. For the rest of the paper we will use $y\sim 10^{-4}$ as
a reference value \cite{Carlstrom}. In the ordinary approach to
the SZE, $\Delta I\left( \nu \right) =y\,f\left( \nu ,T\right) $,
where $f\left( \nu ,T\right) $ is a spectral distortion function
usually computed by means of an approximate solution of the
Kompaneets diffusion equation. The precise form of $f\left( \nu
,T\right) $ is not important for the present paper and the reader
is encouraged to consult the reviews. One interesting fact is that
the change of brightness temperature of the CMBR in the
Rayleigh-Jeans region is simply given by:
\begin{equation}
\frac{\Delta T_{CMB}}{T_{CMB}}=-2y  \label{yxxx}
\end{equation}

Now, the $y$ factor may enter into the SFT approach to the SZE if
we slightly modify Eq. (\ref{Sk1}) so that the distorted spectrum
will now read:
\begin{equation}
I\left( \nu \right) =\int_{0}^{\infty }I_{o}\left( \omega \right)
S(\omega -(1-2y)\,\nu )d\omega \label{buena2}
\end{equation}
Fig. 2 shows that the ``fine tuning'' given by the $y$ parameter
is very effective. Indeed, all the features of the SZE are
reproduced accurately enough to consider SFT as an appropriate
tool to analyze CMBR distortions.
\bigskip
\begin{figure}
\epsfxsize=3.4in \epsfysize=2.6in \epsffile{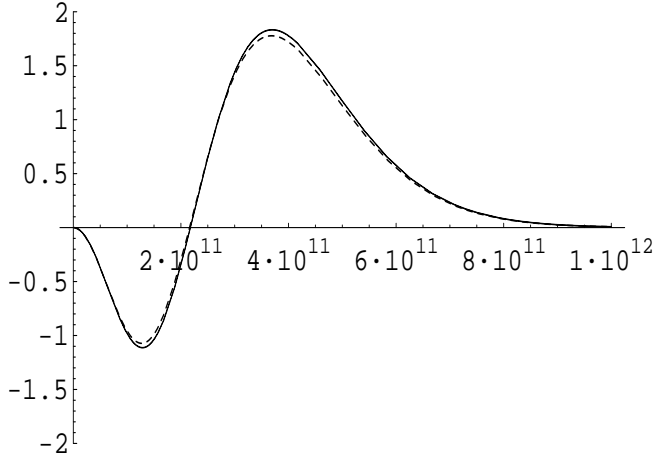}\vspace{0.5cm}
\caption{The same as in Fig. 1, but taking into account Eq.
(\ref{buena2}). The reproduction of the main features of the SZE
curve is clear.}
\end{figure}
\vspace{0.5cm}
\bigskip

\textbf{IV. Discussion}

\bigskip

There is a physical reason that supports the crucial center shift
in the convolution integral (\ref{buena2}). Compton effect
generates a shift in frequency that, integrated along the photon
path in the cluster, roughly indicates that the center of the
structure factor, for a fixed frequency $\nu $, must be displaced
by an amount proportional to the $``y"$ parameter. This picture is
consistent with the description of comptonization in clusters
included in Rybicki and Lightman's text, in which the $``y"$
parameter is defined as a tool to determine whether a photon will
significantly change its energy (frequency) while traversing the
medium \cite {Rybicki}. Thus, $``y"$ is basically the average
fractional frequency change due to Compton scattering, considering
all its flight across the intra-cluster gas. The calculation here
presented \emph{is not} a simple fit. Moreover, if the Compton
factor is reduced by a factor of 10 (corresponding to a
\emph{physical} multiplying factor of 0.325 for $\alpha$ in
Eq.(\ref{alpha1})), the reproduction of the SZE curve by means of
Eq.(\ref{buena2}) is practically perfect.

\bigskip

\begin{figure}
\epsfxsize=3.4in \epsfysize=2.6in \epsffile{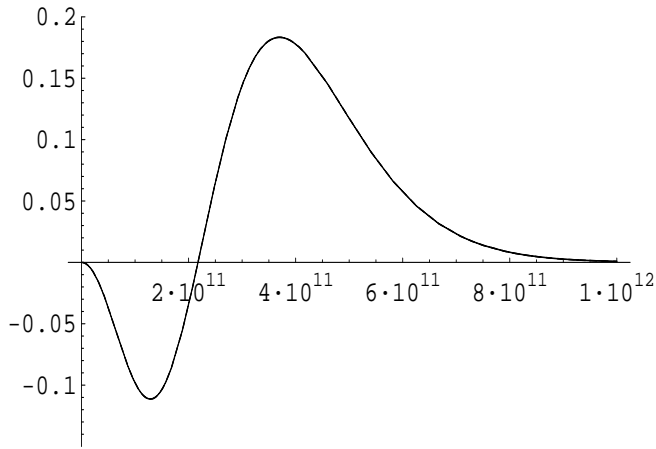}\vspace{0.5cm}
\caption{The same as in Fig. 2, but with the Compton factor
lowered by a factor of 10}
\end{figure}
\vspace{0.5cm}

\bigskip
All parameters included in the basic equation (\ref{buena2}) have
a proper physical meaning and other uses of the formalism in CMBR
related problems seem plausible. To the author's knowledge, this
is the first time that SZE-type curves are obtained by means of
convolution integrals. The basic curve can be associated with a
standard structure factor, while ``fine tuning'' can be related to
the gas comptonization parameter $``y"$. Obviously, the integral
expression (\ref{buena2}) must be related to the standard solution
of the Kompaneets equation that ordinarily describes the
non-relativistic SZE.

\bigskip

The work here presented has established some basics about a SFT
approach to the SZE. The results shown here suggest that structure
factors may provide alternative approaches to CMBR distortion
problems. Since line broadening is also present in all dissipative
systems, proper convolution integrals involving Lorentzian
structure factors may yield alternative approaches to light
scattering in cosmological processes, such as recombination. In
recombination, photon scattering due to acoustic waves must be
taken into account. Other future work includes the relativistic
generalization of this work in terms of a proper Maxwellian.
Finally, a kinetic theory approach to structure factor theory by
means of the techniques suggested in Ref. \cite {Sandoval} will
soon be addressed.

\bigskip

The  author thanks Leopoldo S. Garc\'{\i}a-Colin (UAM-I) and Roy
Maartens (Portsmouth) for valuable comments and suggestions. This
work was supported by a CONACyT postdoctoral grant. The author
also thanks the Relativity and Cosmology Group at Portsmouth for
hosting him while this work was done.

\end{document}